\documentclass[conference]{IEEEtran}
\IEEEoverridecommandlockouts

\usepackage{natbib}
\usepackage{amsmath,amssymb,amsfonts}
\usepackage{algorithmic}
\usepackage{graphicx}
\usepackage{float}
\usepackage{textcomp}
\usepackage{xcolor}
\usepackage{outlines}
\usepackage{graphicx}
\usepackage{subcaption}
\usepackage{hyperref}

\usepackage{xspace}
\usepackage{url}
\usepackage{fancyhdr}
\def\BibTeX{{\rm B\kern-.05em{\sc i\kern-.025em b}\kern-.08em
    T\kern-.1667em\lower.7ex\hbox{E}\kern-.125emX}}

\newcommand{\TODO}[1]{}
\newcommand{\nelson}[1]{}
\newcommand{\shweta}[1]{}
\newcommand{\zhuoran}[1]{}

\newcommand{\FMwork}{FMwork }

\begin{document}

%\title{\FMwork: Benchmarking Meta-Metrics and Best Practices for LLM Inference}
%\title{\FMwork: Best Practices for Inference Benchmarking}
%\title{\FMwork: Best Practices and Meta-Metrics for LLM Inference Benchmarking}
% \title{\FMwork: Meta-Metrics and Best Practices for LLM Inference Benchmarking}
%\title{\FMwork: Best Practices for LLM Inference Benchmarking}
% \title{EffiBench: Meta-Metrics and Best Practices for System-Level Performance Benchmarking of LLM Inference Workloads}
\title{Meta-Metrics and Best Practices for System-Level Inference Performance Benchmarking}

 \author{\IEEEauthorblockN{
     Shweta Salaria, Zhuoran Liu, Nelson Mimura Gonzalez}
     \IEEEauthorblockA{
         IBM Research, Yorktown Heights, New York \\
         \{shweta.salaria, zhuoran.liu, nelson\}@ibm.com}
 }

\maketitle

%\sloppy

\begin{abstract}
Benchmarking inference performance (speed) of Foundation Models such as Large Language Models (LLM) involves navigating a vast experimental landscape to understand the complex interactions between hardware and software components. 
However, evaluating every possible test configuration is impractical, unfeasible and unnecessary. 
To address this challenge, we introduce FMwork, a comprehensive and methodical approach to creating a controlled testing environment that accurately reflects and characterizes performance.
\FMwork comprises a set of benchmkaring best practices with three key components: 1) meta-metrics, 2) parameter selection, and 3) strategic cost-performance evaluation.
Meta-metrics account for time and resources spent on benchmarking and the relative accuracy of the results compared to a larger body of measurements, representing the complete experimental space.
\FMwork operationalizes the meta-metrics and provides efficient strategies for parameter selection and cost-performance analysis.
Using the framework, we show up to 24x improvement (speedup and/or resource savings) running sweeps of experiments compared to the ground truth. 
Even already considering a subset of experiments as reference point (using the power of two for batch sizes), reducing experimental output size from 1024 to 128 tokens yields another 2.7x gain while keeping 96.6\% accuracy for an evaluation using Llama 3.1 8B model.
\end{abstract}

\section{Introduction}

% General introduction about LLMs

Foundation Models such as Large Language Models (LLMs) are transforming the field of Artificial Intelligence (AI), specifically in the domain of Natural Language Processing (NLP). These deep neural networks are trained on vast corpora of text data~\cite{raffel2020exploring, bommasani2021opportunities}, enabling them to understand and generate human-like text in various contexts and applications. LLMs function by processing sequences of tokens (words) and predicting what comes next based on patterns learned during training. The development of LLMs has progressed rapidly, starting with GPT~\cite{brown2020language, radford2019language} in 2018 and continuing with today’s advanced models like GPT-4~\cite{achiam2023gpt}, Claude~\cite{anthropic2023claude}, Llama~\cite{touvron2023llama,touvron2023llama2,grattafiori2024llama}, Granite~\cite{granite2024granite}, and others. Each new generation demonstrates increasingly improved abilities to understand context, follow instructions, reason through problems, and generate coherent, relevant text.

% Basics about LLM inference

LLM inference uses a trained language model to generate responses to inputs. Unlike the training phase, which requires enormous computational resources and typically runs once, inference happens repeatedly in real-time applications such as chatbots, code assistants, and search engines. These use cases introduce technical challenges across several dimensions~\cite{kwon2023efficient, NvidiaLLMInferenceBlog}: computational efficiency (balancing output quality with speed and resource consumption), latency control (delivering responses quickly enough for seamless user experiences), memory management (accessing billions of parameters efficiently during operation)~\cite{kwon2023efficient,rajbhandari2020zero}, and decoding methodology~\cite{vaswani2017attention, pope2023efficiently} (selecting generation strategies that produce appropriate, diverse outputs). These challenges shape how language models perform in real-world settings and determine their practical utility across different deployment environments.

% Hardware and software advancement to support LLMs

As LLMs continue to evolve in size and complexity, there has been a swift advancement in both the software ecosystem and hardware needed to support these massive models. Hardware innovations have focused on increasing computational efficiency and reducing latency. The rise of specialized AI accelerators, such as GPUs and TPUs~\cite{emani2022comprehensive, wang2019supporting}, is driven by the increasing parallel processing demands of LLMs. Memory improvements, including High Bandwidth Memory (HBM) and unified memory architectures, aim to reduce latency by improving data transfer and accessibility. Modern hardware increasingly supports low-precision computation (e.g., INT8, FP16), enabling model quantization and faster, energy-efficient inference \cite{lin2024awq, dettmers2023case} with minimal loss in accuracy. Parallel computing architectures allow model and tensor parallelism~\cite{shoeybi2019megatron,rasley2020deepspeed,li2024efficient}, enabling inference with very large models by distributing computation across multiple processors~\cite{huang2019gpipe,li2023alpaserve}. Software advancements complement hardware by optimizing how models are executed. Inference-optimized software frameworks such as TensorRT~\cite{nvidia-trtllm}, ONNX Runtime, and DeepSpeed-Inference~\cite{rasley2020deepspeed} enhance performance through hardware-aware compilation and optimization techniques~\cite{rasley2020deepspeed, nvidia-trtllm}. Quantization, pruning, and model distillation reduce model size and computation without significantly affecting output quality. Techniques like KV caching~\cite{kwon2023efficient} and FlashAttention~\cite{dao2023flashattention, shah2024flashattention} minimize redundant computation during token generation. Together, these developments are not only accelerating LLM inference but also redefining how benchmarking is conducted, with more emphasis on end-to-end pipeline efficiency, energy consumption, and cost-effectiveness in real-world deployment scenarios. In this rapidly evolving space, understanding the capabilities and limitations of new hardware platforms coupled with software improvements is essential for accurately benchmarking LLM inference performance and for making informed decisions about infrastructure investments and model deployment strategies.
%Moreover, software and system-level optimizations such as kernel fusion, dynamic batching, quantization-aware inference, and KV cache reuse are being tightly coupled with new hardware to push the limits of performance further. 

% Challenges w.r.t inference benchmarking 

Benchmarking inference performance of large language models is a multifaceted challenge influenced by technical, operational, and evaluation complexities. This process is further complicated by the vast number of possible configurations that can significantly affect results. One of the core tensions lies in the trade-off between latency and throughput. Applications such as conversational agents prioritize low latency for real-time interaction, while tasks like large-scale document processing benefit more from high throughput. Batch size (concurrency) plays a critical role. While larger batch sizes improve throughput by maximizing hardware utilization but also introduce higher latencies for individual requests. Since the optimal batch size depends on the model architecture, deployment hardware, and application requirements, benchmarking should evaluate performance across a range of batch sizes rather than relying on a single setting. Another major challenge is the selection of evaluation metrics. Common metrics include tokens per second, inter-token latency, cost per million tokens, and energy consumption. However, their relevance varies depending on the use case. As a result, benchmarking LLM inference requires a careful, context-aware approach~\cite{reddi2020mlperf, agrawal2024vidur, emani2022comprehensive} that considers trade-offs, configuration diversity, and performance metrics.

% Define problem statement - need for a benchmarking methodology

A key challenge in benchmarking LLM inference is the lack of standardized benchmarking practices. While efforts like MLPerf~\cite{reddi2020mlperf} and Hugging Face’s Open LLM Leaderboard~\cite{open-llm-leaderboard} are valuable, inconsistencies in environments, generation settings, and even tokenization approaches make it difficult to compare models directly. So, the challenge for us was to answer the question: \textit{"Can we develop a benchmarking methodology for LLM inference workloads that strategically selects representative configurations from the vast, multidimensional parameter space—enabling evaluation that is both time-efficient and cost-effective?".}

%Overall, benchmarking LLMs involves navigating the challenges of extreme computational demands, hardware limitations, and the probabilistic outputs, all of which must be considered when creating fair and accurate performance metrics. This makes it more challenging to directly compare different models, systems, or configurations, especially as the scale of LLMs continues to increase.

% Our approach to inference benchmarking

To answer the above question, we propose a systematic methodology that reduces the size and complexity of the experimental space while maintaining fidelity to real-world performance. This method focuses on identifying a representative set of configurations that reflect typical deployment scenarios across diverse application domains. By narrowing the benchmarking scope to high-impact parameter combinations, we can produce actionable insights with reduced time and computational cost. Furthermore, the proposed approach is designed to scale effectively across multi-node cluster environments.
%This targeted approach not only simplifies benchmarking for developers and researchers but also increases reproducibility and relevance. 

Our contributions are:

\begin{enumerate}
    \item We introduce FMwork\footnote{https://github.com/IBM/fmwork}, an open-source, comprehensive and efficient approach to evaluate the inference performance of large language models based on inputs (model, input size, output size, batch size, precision etc.), backends (vLLM, TensorRT-LLM etc.) and outputs (time, cost, power etc.).
    \item We present meta-metrics to aid the design of a representative test setup by quantifying time, cost, and resources spent on experimentation and comparing accuracy to the complete sample space.
    \item We thoroughly discuss \FMwork operationalizing the meta-metrics through parameter selection strategies that optimize sampling efficiency, cost-performance analysis that quantifies resource requirements, and synthetic input, model generation that ensures representativeness and flexible evaluation.
    \item The results show up to 24x improvement efficiency as compared to ground truth datasets; reducing the experimental output size from 1024 to 128 tokens yields an additional 2.7x gain while preserving 96.6\% accuracy.
\end{enumerate}

This paper is organized as follows.
Section~\ref{sec:background} presents background and our motivation for defining our framework;
Section~\ref{sec:metametrics} proposes our meta-metrics to methodically define the exploration space to be considered during benchmarking;
Section~\ref{sec:design} presents our design and evaluation of the framework.
Section~\ref{sec:relatedwork} presents related work.
Finally, Section~\ref{sec:conclusion} concludes our paper.

\section{Background \& Motivation}
\label{sec:background}

% ------------------------------------------------------------------------------

%Inference benchmarking presents significant challenges due to the complex interactions between hardware and software components~\cite{NvidiaLLMInferenceBlog,kwon2023efficient,rasley2020deepspeed,nvidia-trtllm}, as well as parameter explosion~\cite{brown2020language,wolf2020transformers,rajbhandari2020zero}. 
This section outlines the principle components of inference benchmarking and provides the rationale for establishing an efficient benchmarking methodology.

%This section outlines the key dimension of these challenges and motivates the definition of an efficient benchmarking methodology.

% ------------------------------------------------------------------------------

\subsection{Benchmarking Parameters}

Inference performance is governed by several parameters:
\begin{itemize}
    \item \textbf{Hardware} -- where we are running. 
    The device may be a CPU (specified by model, number of sockets per node, cores per socket, threads per core); a GPU (defined by its model, architecture, and available memory); or another type of specialized accelerator. Depending on the model’s size and the hardware characteristics, different parallelism strategies such as tensor parallelism, data parallelism can be used to fit the model. Devices with larger memory capacity may support greater batch sizes (i.e., higher concurrency); however, this typically comes at the cost of increased latency. Therefore, sufficient computational power is required to sustain the resulting workload efficiently.
    %Several aspects of both workload and hardware affect balancing aspects such as this~\cite{agrawal2024vidur,emani2022comprehensive,lazuka2024llm}.
%
    \item \textbf{Model} -- what we are running. 
    This is the object being benchmarked — for instance, an LLM. It typically encompasses a particular model architecture and number of parameters. Larger models (i.e., with more parameters) require more memory and resources to run than smaller models. The same model family may change internal architecture over time~\cite{brown2020language,touvron2023llama,touvron2023llama2,grattafiori2024llama}. Architectures may evolve as well to incorporate technological improvements (affecting both quality and speed)~\cite{ainslie2023gqa,vaswani2017attention,dao2023flashattention, shah2024flashattention}. There is also the distinction between dense models, such as llama-3.1-8b~\cite{grattafiori2024llama}, and mixture of experts models (MoE), such as mixtral-8x7b~\cite{jiang2024mixtral} and llama-4-maverick~\cite{meta2025llama4}, DeepSeek-V2~\cite{liu2024deepseek}.
\end{itemize}

The next aspect is \textit{workload} -- how we are running:
\begin{itemize}
    \item \textbf{Input size}:
    Also referred to as prompt length or sequence length, this parameter represents the number of tokens in the input prompts. It is one dimension of the model's input tensor during inference operations.
    \item \textbf{Batch size}:
    This parameter defines the number of sequences processed in parallel, forming the other dimension of the input tensor during inference operations. In deployment / server-mode scenarios, batch size is correlated to the number of concurrent users an inference server is simultaneously serving.
    \item \textbf{Output size}:
    The number of new tokens generated in response to the input. Each token generation involves processing the attention mechanism with the previously generated context~ \cite{vaswani2017attention}.
    \item \textbf{Parallelism}:
    This parameter determines how a model is distributed across multiple resources. Due to resource constraints and/or performance requirements, a model can be broken down into smaller parts that can run on multiple devices. Several techniques can be employed to run a model over distributed resources, such as data parallelism~ \cite{li2023alpaserve}, pipeline parallelism~ \cite{huang2019gpipe}, tensor parallelism~ \cite{shoeybi2019megatron,rasley2020deepspeed,li2024efficient}, expert parallelism~ \cite{zhu2025megascale,rajbhandari2022deepspeed}, and hybrid parallelism~ \cite{narayanan2021efficient}.
    \item \textbf{Precision}:
    This is how many bits or bytes are used to represent numbers across the execution of a model. Several aspects can be controlled with regards to precision of, for instance, an LLM.
    \begin{itemize}
        \item First, model \textbf{weights} can be distributed and then loaded in float32, float16, bfloat16, float8, int8, and int4, to name a few. The process of reducing the number of bits used to represent numbers is typically referred to as quantization~ \cite{lin2024awq}. A model might be trained and distributed in bfloat16 precision but then loaded into an inference engine in float8 by means of dynamic quantization~ \cite{dettmers2023case}. Models can be statically quantized as well~ \cite{lang2024comprehensive}, so that it is not necessary to spend time during execution to quantize weights (therefore improving overall performance). Quantization in this case can help fitting a large model into smaller devices (i.e., devices with less memory) by reducing its memory footprint. However, quantization can affect model accuracy.
        \item After weights, the next aspect that can be controlled is \textbf{activations}, which are the values computed as data flows through the model's layers~ \cite{dao2023flashattention,wang2025exploring}. This affects both memory usage and computational requirements, but quantized activations can be more sensitive to precision loss than weights~ \cite{hubara2018quantized}.
        \item Finally, the \textbf{kv-cache} can also be quantized. The key-value cache stores intermediate attention computations to avoid redundant calculations during autoregressive generation. For long-context inference, kv-cache can dominate memory usage~ \cite{kwon2023efficient}; consequently, reducing precision may affect how long a context can be when running a particular model.
    \end{itemize}
\end{itemize}

For comprehensive inference performance characterization, these parameters must be systematically explored across their operational ranges. Similar to observations in MLPerf benchmarks, LLM inference scenarios encompass significant variability in both batch sizes and input lengths -- from single-query processing with strict latency constraints to high-throughput batch processing with variable sequence lengths. This parameter space complexity suggests the need for a methodologically sound approach to ensure representative coverage while maintaining experimental tractability, particularly as each parameter independently influences multiple performance metrics across different deployment contexts.

% ------------------------------------------------------------------------------

\subsection{Performance Metrics}
LLM inference operates in two distinct computational phases~\cite{NvidiaLLMInferenceBlog, kwon2023efficient}: the \textbf{prefill} phase processes the entire input sequence at once with costs scaling with input length, while the \textbf{decode} phase processes one new token at a time using kv-cache to avoid redundant computations. Based on these differences, we define three key metrics:
\begin{itemize}
    \item \textbf{Time to First Token (TTFT)}: This metric measures the duration from request initiation to first token generation, primarily reflecting prefill phase performance. TTFT indicates system responsiveness in interactive applications where low latency impacts user experience.

    \item \textbf{Inter-token Latency (ITL)}: This metric measures the average time interval between consecutive token generations, calculated by tracking time differences between adjacent tokens. ITL primarily reflects decode phase performance and is critical for streaming applications where inconsistent generation speeds can cause noticeable output stuttering.

    \item \textbf{Throughput (THP)}: This metric measures token generation capacity across concurrent requests, calculated by dividing total tokens produced (a function of batch size and per-sequence output size) by generation time (excluding TTFT). Measured in tokens per second, THP reflects system efficiency during steady-state operation.
\end{itemize}
Performance metrics exhibit interdependence, creating trade-offs where optimizing throughput typically sacrifices latency and vice versa. This necessitates multi-objective evaluation approaches, while diverse LLM implementations introduce additional variability that comprehensive benchmarking must address.

% ------------------------------------------------------------------------------

\subsection{Multidimensional Complexity in LLM Ecosystems}
Beyond control parameters and performance metrics, several additional dimensions exponentially expand the LLM benchmarking space:
\begin{itemize}
    \item \textbf{Model Architectures:} Transformer-based models~\cite{vaswani2017attention} exhibit fundamental operational differences: encoder-only models like BERT~\cite{devlin2019bert} process input sequences in parallel; decoder-only models operate in sequential prefill and decode phases~\cite{pope2023efficiently}; and encoder-decoder architectures~\cite{lewis2019bart, raffel2020exploring} combine both approaches. These architectural variations require independent characterization, multiplying the overall benchmarking challenge.
    
    \item \textbf{Scale and Evolution:} Language models vary dramatically in size~\cite{bommasani2021opportunities, wolf2020transformers}, with parameter counts from millions to billions~\cite{brown2020language, chowdhery2023palm}, necessitating specialized optimization strategies~\cite{wang2019supporting, narayanan2021efficient, rasley2020deepspeed, rajbhandari2020zero}. Rapid architectural evolution—exemplified by Llama's progression from standard multi-head attention to grouped-query attention (GQA)~\cite{ainslie2023gqa}—renders comprehensive evaluations quickly obsolete, making exhaustive testing impractical.
    
    \item \textbf{Inference Frameworks:} Specialized serving systems like vLLM~\cite{kwon2023efficient}, TensorRT-LLM, Text Generation Inference~\cite{huggingface-tgi, iglesias2024llm}, and PyTorch~\cite{paszke2019pytorch} implement distinct optimization strategies. While these frameworks report performance metrics, they lack systematic analysis methods or efficient experimental methodologies, creating additional dimensions where each framework-model combination exhibits unique performance characteristics.
\end{itemize}.

% ------------------------------------------------------------------------------

\subsection{The Need for Efficient Benchmarking Methodology}
The convergence of multiple factors—control parameters, interdependent metrics, diverse architectures, rapid evolution, and proliferating frameworks—creates an overwhelmingly complex experimental space for LLM benchmarking. Traditional exhaustive testing becomes intractable, yet organizations require reliable performance insights for deployment decisions. This challenge necessitates \FMwork, a principled methodology delivering representative performance characterization while drastically reducing experimental costs.

\FMwork quantifies the trade-off between experimental investment and representational fidelity through strategic parameter selection and systematic sampling techniques. The framework provides scientifically grounded protocols that deliver actionable insights with minimal overhead, meeting the urgent industry demand for efficient yet reliable performance evaluation methodologies.
% ==============================================================================

\section{Meta-Metrics}
\label{sec:metametrics}

Benchmarking is a methodical approach to sampling and characterizing a virtually infinite experimental space. This process involves strategically selecting representative points within a vast multidimensional parameter landscape to generate meaningful insights about system behavior and performance characteristics. The core challenge of effective benchmarking lies in determining which regions of this expansive space to explore, as exhaustive evaluation is costly and computationally intractable. Through careful experimental design and principled sampling strategies, benchmarking enables researchers to make generalizable assertions about system capabilities, performance trade-offs, and optimization opportunities across diverse operational scenarios that would otherwise remain unexplored.

In that sense, inference benchmarking for large language models introduces considerable complexity. For any given evaluation, researchers must specify a particular model architecture, precision format (e.g., 16-bit, 8-bit, other quantization options), parallelism configuration (e.g., tensor parallelism, data parallelism, expert parallelism), and inference backend (e.g., Transformers, vLLM, TRT-LLM). Within this configuration framework, three essential parameters fundamentally shape inference behavior: input sequence length, output generation length, and batch size. Modern models can process inputs and generate outputs ranging from single tokens to 128k (128x1024 = 131,072) tokens, while batch sizes may vary from individual requests to batches of 1,024 or more concurrent prompts. This creates an experimental space with trillions of potential measurement points for a single model-precision-parallelism-backend combination. When multiplied across the ecosystem of available models, precision options, parallelism strategies, and inference frameworks, exhaustive benchmarking becomes (even more) prohibitively expensive. A complete characterization of even a subset of this vast parameter landscape would require computational resources beyond what is typically available for benchmarking exercises or what is even natural or realistic. A simple calculation: if each experiment took just 20 seconds to run (which is extremely optimistic); if we had 1 million devices (e.g., high-power GPUs) to execute such experiments, it would still take over 5 years to finish all sweeps. If each card consumed 700W, that represents a 700 MW cluster. For reference, the largest supercomputers in the world currently consume around 20-30 MW; so that would represent over 20 modern supercomputers. To finish such a sweep in three months (assuming that anything longer than that would yield the sweep obsolete before it even finished) would require an infrastructure of almost 16 GW. For reference, the largest power plants in the world produce around 20 GW or less.

Reducing either of these dimensions by a few orders of magnitude could bring total time and cost down to reasonable numbers. This suggests we can make a decision about which experiments to run and therefore 1) we could calculate how much time these experiments would take and how much they would cost; and 2) we could assess, at the very least from a relative perspective, how close we are to the complete sample space. The next subsections explore these concepts by elaborating on metrics to quantify time (cost) and accuracy; and then provide guidelines on how to select parameters based on these metrics while making simplifications that favor familiar patterns, such as sequences and multiples, which enhance human interpretability even if they do not perfectly optimize mathematical representation of the sample space. Finally, we explore different methods to evaluate and compare cost-performance for different datasets.

The first metric we consider in our analysis is \textbf{time} spent to run experiments. This can be measured and, if necessary, projected. We can then calculate the cost to run the experiments by multiplying time by resource cost per time. The cost of running all experiments for a particular set of input sizes $I$, output sizes $O$, and batch sizes $B$, is:
\begin{equation}
C = C_{unit} \sum_{i \in I} \sum_{o \in O} \sum_{b \in B} t(i, o, b)
\end{equation}
where
$C_{unit}$ is device(s) cost per unit time; 
$t(i,o,b)$ is the time required to run an experiment with input size $i$, output size $o$, and batch size $b$;
and the triple summation iterates through all combinations of parameter values in the specified sets.
Time will vary with all inference parameters, including the hardware being benchmarked. Cost per time may vary with hardware as well. An experimental sweep might require covering several hardware devices of interest. Note also the number of devices does not affect the calculations; more device will only lead to a shorter end-to-end benchmarking time (which is certainly desirable).

The second metric we consider is \textbf{accuracy} — representational fidelity, as in how well the selected parameters and derived results capture the larger experimental space. This creates a natural tension with the first metric; accuracy might be improved at the cost of additional time and resources. We define a set $G$ of combinations of input sizes $I_G$, output sizes $O_G$, and batch sizes $B_G$ that we adopt as ground truth; i.e., representing a relatively large subset of the complete experimental space. We then define a set $M$ of measured combinations; i.e., a set of combinations of input $I_M$, output $O_M$ and batch $B_M$ sizes that are executed as experiments on real devices. Based on $M$ we determine $P$ — a superset of $M$ that uses different techniques (e.g., interpolation, extrapolation) to project or "fill the gaps" compared to $G$. For each combination of input size $i$, output size $o$ and batch size $b$ we can calculate the normalized / relative difference between $G$ and $P$, for each point:
\begin{equation}
\delta_{\text{normalized}} = \frac{|f_G(i, o, b)| - |f_P(i, o, b)|}{|f_G(i, o, b) + f_P(i, o, b)|}
\end{equation}
where $f(i, o, b)$ is a metric measured during the experiment (e.g., time to first token, inter-token latency). Based on these numbers we can calculate a global accuracy factor $\Delta$ by taking multiple statistical metrics, such as average, median, percentiles and root mean square error. Finally, we can calculate accuracy-cost factors using different techniques to characterize a particular choice of $I$, $O$ and $B$. A simple efficiency metric $E$ can be defined as:
\begin{equation}
\label{eq:eff}
E(G,P) = \frac{1 - \Delta(G,P)}{C_P / C_G}
\end{equation}
% %
% and a harmonic version as:
% %
% \begin{equation}
% H(G,P) = \frac{2 \cdot (1 - \Delta(G,P)) \cdot (C_G / C_P)}{(1 - \Delta(G,P)) + (C_G / C_P)}
% \end{equation}
% %
% Both will severely penalize accuracy losses; the harmonic mean will be slightly more forgiving.

\section{Design \& Evaluation}
\label{sec:design}

% \nelson{
% \centering ------------------------------------------------------------
% 
% Thoughts:
% 
% \begin{outline}
%     \1 More and more I think a lot of this is \textbf{[Method]}; then the meta-metrics part is \textbf{[Discuss]}. In \textbf{[Method]} we show all our tricks; HOW they work; in \textbf{[Discuss]} we show WHY they work.
% \end{outline}
% }
% \nelson{
% \begin{outline}[enumerate]
% \textbf{[Method]} Benchmarking basics
%         \1 Parameters
%         \1 Results (outputs, metrics, etc)
%         \1 Encoders vs decoders; prefill vs decoding;
% \end{outline}
% }

Our benchmarking methodology provides a systematic approach to evaluate LLM inference performance while judiciously managing experimental complexity. By implementing the meta-metrics principles introduced in Section~\ref{sec:metametrics}, we balance efficiency with representational fidelity, maintaining experimental accuracy while significantly reducing computational requirements. We identify the critical variables that significantly impact inference behavior and develop principled strategies to sample the vast experimental space efficiently without sacrificing analytical depth. Through controlled exploration of the interactions between model architectures, hardware configurations, and workload characteristics, \FMwork enables comprehensive performance analysis across diverse operational conditions.
\subsection{Overview}
\begin{figure}[t]
    \centering
    \includegraphics[width=0.9\linewidth]{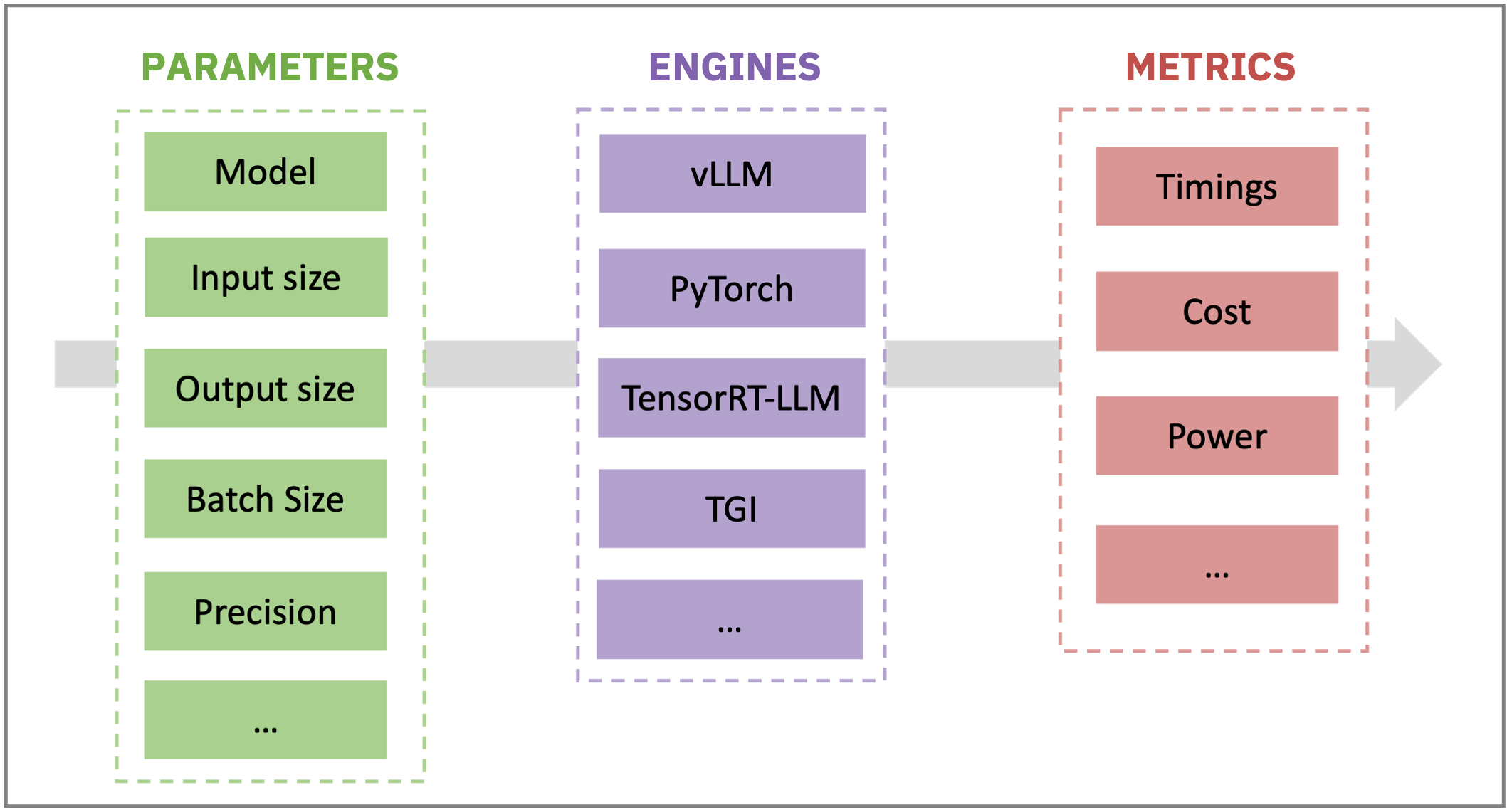}
    \caption{\FMwork: A structured approach to LLM inference benchmarking organized around three key components: (1) Inputs—control parameters that can be systematically varied prior to model execution, including model architecture, input sequence length, output size, batch size, and precision; (2) Backends—inference systems that execute the models; and (3) Outputs—performance metrics obtained after benchmarking, including timing measurements, cost, and power efficiency.}
    \label{fig:benchmarking-framework}
\end{figure}

As illustrated in Figure~\ref{fig:benchmarking-framework}, \FMwork implements a tripartite structure that methodically examines the relationship between input parameters, inference backends (such as vLLM, PyTorch, TensorRT-LLM, and TGI), and resulting performance metrics. This architecture operationalizes the meta-metrics established earlier, providing a concrete framework to implement the cost-accuracy optimization principles we developed. By organizing our benchmarking methodology around these three interconnected components, \FMwork creates a structured evaluation approach that allows us to strategically sample the vast parameter space. This methodology enables us to precisely quantify experimental time requirements while maintaining high representational accuracy of the complete sample space.

To operationalize this approach, we strategically select parameter combinations based on their impact on both experimental time and representational accuracy. In the following sections, we methodically examine how our framework implements the Meta-Metrics methodology through parameter selection strategies that optimize sampling efficiency, cost-performance analysis that quantifies resource requirements, synthetic inputs generation that ensures workload representativeness, and synthetic model generation that enables flexible evaluation across different model architectures, hardware configurations, and inference backends.

\subsection{Parameter Selection}
\label{sec:discussion/param}

We analyze a series of sweeps covering different parts of the experimental space in different granularities. All tests assume a llama-3.1-8b model operating at BF16 precision, running on an NVIDIA H100 GPU (single GPU, TP-1) on vLLM (version 0.7.2) with single scheduler step in eager mode. We start our analysis with a sweep varying input sizes starting at 128 and incrementing in multiples of 2. We fix output sizes to 1 (to collect TTFT) plus three options: 16, 128, 1024. We vary batch sizes from 1 to 128 in increments of 1 (i.e., maximum granularity). With the resulting dataset $G$, we create different datasets $P$ — wherein we take a subset of batch sizes from $G$ and interpolate the other values using simple linear regression. In other words — if the selected batch sizes include 2 and 4, then for batch size 3 we interpolate both TTFT and ITL using numbers from batch sizes 2 and 4.

\begin{figure}[!ht]
    \centering
    \includegraphics[width=0.900\linewidth]{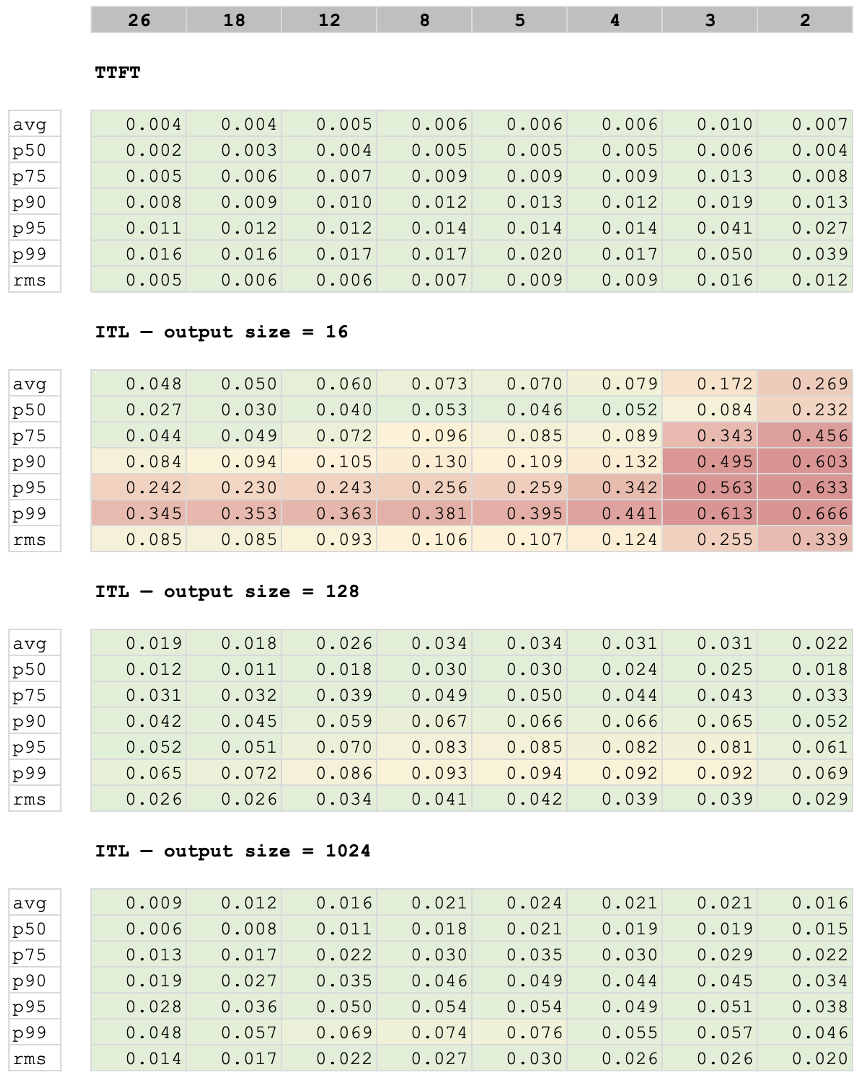}
    \caption{Colored cells indicate relative difference (error; the lower, the better) between $G$ and $P$ subsets projecting TTFT (independent of output size) and ITL (for each output size considered in the sweep). Color coding is green up to 5\% difference; transitioning to yellow at 10\%; and red at 50\% and beyond. Covering input size of 1024 only. Columns (26 to 2) indicate how many batch size values were selected to calculate the projections. The extremes were always selected (batch sizes 1 and 128). Column with value 8 indicates batch sizes as powers of 2.}
    \label{fig:sweep-a1}
\end{figure}

Figure~\ref{fig:sweep-a1} shows different statistics calculated based on the relative difference between measured and projected values when comparing $G$ and $P$, for each subset $P$ (columns in the table), across all batch sizes evaluated. The objective of the experiment is twofold: (1) determine what subset of batch sizes still provides a reasonable number of data points to cover the experimental space; and (2) provide insight on what output size leads to acceptable statistical quality. TTFT projections are independent of output size. ITL projections are calculated for each output size. The columns represent different subsets — the number on the top is the number of points selected. Option "8", for instance, is 1 to 128 in multiples of 2. The results show that output size of 128 is enough to characterize the experimental space while keeping accuracy within 10\%. If we take the average to define our accuracy metric, we achieve 3.4\% of global relative difference with output size 128 and a subset of powers of 2 for batch size. We can even achieve similar or better accuracy with coarser subsets. The same subset of powers of 2 can achieve 0.6\% global difference for TTFT projections.

Regarding time and resource utilization -- the total time to run experiments for all batch sizes with output size 1024 is 2499~seconds. Considering only powers of 2 for batch sizes, output size 1024, the total time is 111~s. Using output size 128, total time is 19~s. From Figure~\ref{fig:sweep-a1}, the average error for ITL, column = 8 values, output size 1024 is 0.021 and for output size 128 is 0.034. The efficiency factor, as defined in Equation~\ref{eq:eff}, can be calculated by multiplying the speedup by one minus the error — so, $(2499/111)\times(1-0.021)=22$ for output size 1024; and $(2499/19)\times(1-0.034)=127$ for output size 128. In other words -- using powers of 2 and output size 128 can save almost 130x in time and resources. \zhuoran{This represents our most dramatic efficiency case—a 130x improvement that showcases the upper bound of what our parameter selection strategy can achieve.} For the same set of batch sizes, comparing output size 1024 and 128, running experiments with shorter outputs saves almost 6x time and resources without significant accuracy loss.

If we expand this analysis to all other input sizes sweeped, total time for all batch sizes is 32270~s at output size 1024 (and 12675~s at output size 128). Using powers of 2 for batch sizes the total times are 3483~s and 1301~s. The error factors (not in the figure) are 0.034 and 0.045, respectively. If we do the same calculation that was presented previously, the efficiency factors across all input sizes are, therefore, 9x for output size 1024 and 24x for output size 128. \zhuoran{This 24x efficiency factor demonstrates the scalability of our approach across diverse input sizes.} Looking only at batch sizes in powers of 2, comparing output size 1024 vs 128, the speedup factor is 2.7x. \zhuoran{This additional 2.7x gain from reducing output size offers significant benefits with minimal accuracy impact.}

% ==============================================================================

\subsection{Cost-Performance Analysis}
\label{sec:discussion/cost-perf}

Performance (as in speed) is a key aspect when comparing models and hardware options. Accuracy certainly is another key aspect -- out of the scope of our framework, and potentially independent of hardware option. However, resource utilization is a third key aspect of a comprehensive analysis. More powerful hardware options will naturally perform better than lighter ones, but when we normalize performance by, e.g., power utilization or cost in general, different trends might appear.

\begin{figure}[!h]
    \centering
    \includegraphics[width=1.0\linewidth]{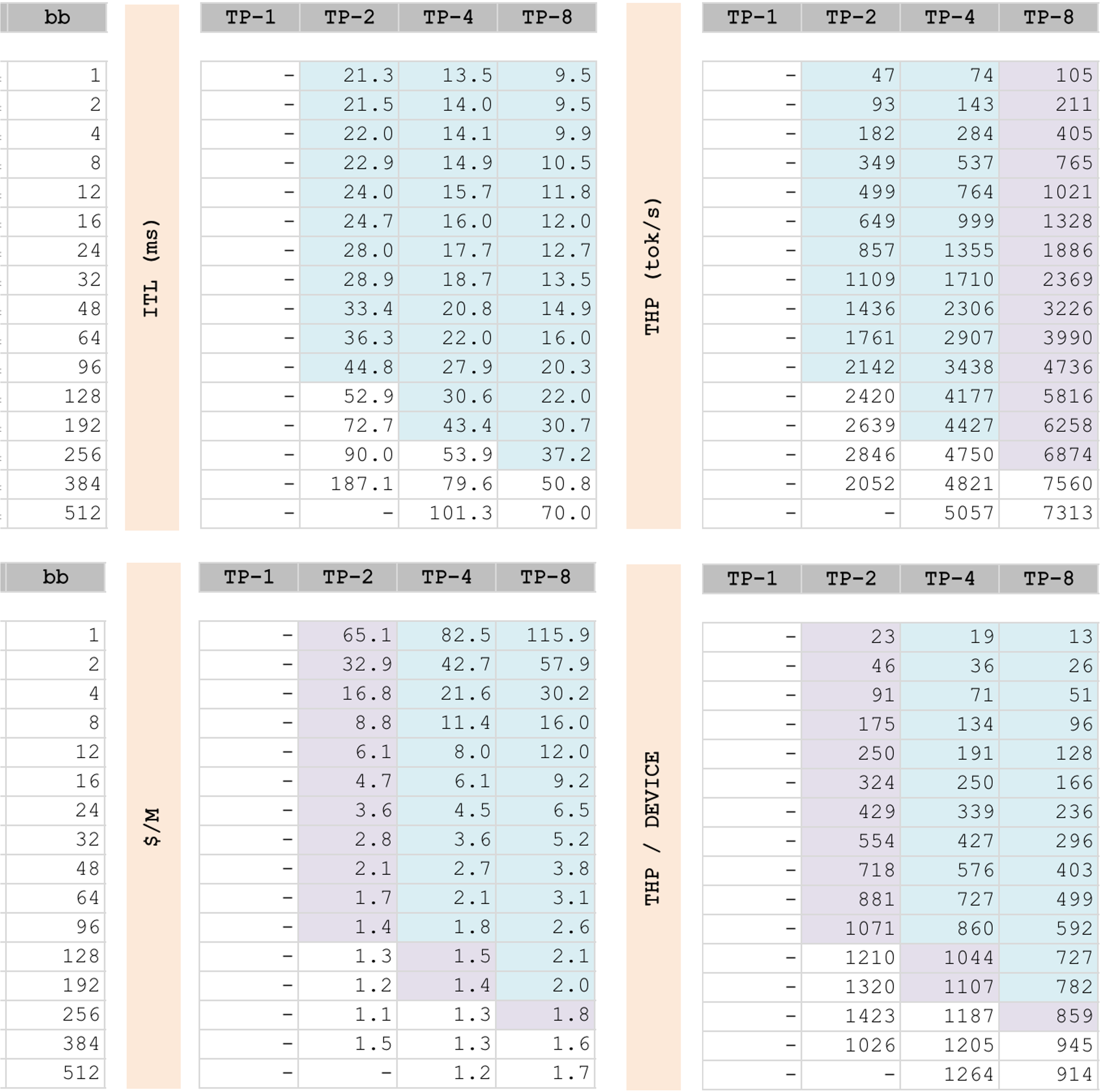}
    \caption{Cost-performance analysis of llama-3.3-70b. Tables show inter-token latency (\texttt{ITL}, in milliseconds), throughput (\texttt{THP}, in tokens per second), throughput normalized per device (\texttt{THP/DEVICE}), and cost per million tokens (\texttt{\$/M}, in dollars) varying batch size (bb i.e. concurrency level) for llama-3.3-70b (16-bit precision).}
    \label{fig:cost-perf}
\end{figure}

Figure~\ref{fig:cost-perf} shows the results of a cost-performance analysis done using data from llama-3.3-70b runs (16-bit precision). We executed the experiments using NVIDIA H100 cards varying number of cards (tensor parallelism) used on each experiment -- from TP-1 up to TP-8. Since the model requires around 140 GB of GPU memory just to load the weights, single card runs do not work (each H100 has 80 GB of memory). Experiments were executed using vLLM version 0.8.4 (engine version = 0, num. scheduler steps = 8). 

There is a lot to unpack in the figure. We present four tables reflecting the results of experiments with input size (sequence length) of 1024 tokens varying batch size (concurrent sequences) from 1 up to 512, when applicable. The top left table shows \texttt{ITL} or inter-token latency, in milliseconds. As expected, as we increase the number of sequences processed in parallel, token latency increases. In contrast, the top right table shows \texttt{THP} or throughput, in tokens per second. Also as expected -- as we increase the number of sequences being processed in parallel, the overall throughput increases. Combining these two pieces of information: as we increase batch size, each individual sequence might see some performance degradation (i.e., it takes longer to generate the next token) but the overall performance of the card (or the system) improves, as we are able to process more sequences at a time. Of course, at some point we saturate the GPU -- in which case, throughput ceases to increase and latency essentially increases linearly with batch size.

If we are running an application or service whose objective is to process sequences as fast as possible, without any other constraints, then the highest throughput option is desirable. For instance -- for a 4xH100 setup, it would make sense to run at a batch size as large as 512 parallel sequences. That is true for batching systems or batched requests -- but what if we are running a real-time service, like a chat. In this case, we need to establish a \textbf{latency threshold}. This threshold must account for how fast, ideally, tokens should be displayed to the end user. In this analysis we use a latency threshold of 50 ms/tok, which should be a good reference for real-time interaction with LLMs while accounting for the model size~\cite{bommasani2021opportunities, wolf2020transformers}. We highlight cells in blue that reflect results that are within the defined threshold. In that case, for a 4xH100 setup, based on the experimental data we could serve as many as 192 concurrent users from one deployment (i.e., one inference server running on 4xH100 GPUs), assuming an operational context length of 1024 tokens. Different context sizes will have different performance behaviors.

We can also compare different hardware options -- either device count or device type. A typical question that may arise from running any model on any environment is -- how many GPUs should we use to run a particular model? If we looked purely at throughput at a particular batch size, then naturally the larger deployment would be the correct answer. However, if a node has, for instance, 8 GPUs, and a model such as llama 70b can be deployed using 2/4/8 GPUs, then we could deploy the model using 4x2xH100, 2x4xH100, or 1x8xH100 setups. It makes sense therefore to normalize the throughput by number of cards, which is reflected in the bottom right table. On both \texttt{THP} and \texttt{THP/DEVICE} tables, the purple cells indicate the best value across different options, for each batch size. TP-8 naturally is better on \texttt{THP} for all batch sizes, but the \texttt{THP/DEVICE} table shows that for batch sizes up to 96, TP-2 is actually better; then TP-4 for batch sizes 128 and 192; and only then TP-8 outperforms the other options. We could also compute the throughput per node, by multiplying \texttt{THP/DEVICE} by the number of devices in a node, which would lead us to a similar conclusion. So depending on the degree of parallelism expected for a service, it might make sense to operate with different parallelism configurations, or even a mix of configurations to serve different types of demands.

Finally, the bottom left table \texttt{\$/M} shows our cost-performance metric -- dollars per million tokens. This represents how much it costs to generate one million tokens given a particular throughput value. With the throughput number we can calculate how long it would take to produce one million tokens -- then we can combine this number with a cost (or price) number per time (e.g., per hour). We then factor in the number of devices needed to achieve such throughput, leading to the final cost calculation. The numbers in the table were calculated assuming an hourly price of \$5.50 for one H100. Based on the results we observe that, first, cost varies greatly with the concurrency level achieved (e.g., by a service) -- since it follows directly the total throughput that can be reached. This also allows us to compare different hardware options. Again, the results follow essentially what is already observed in the \texttt{THP/DEVICE} table -- but based on the \texttt{\$/M} data, considering the context size being experimented with, it would make sense to serve this model either using two or four cards. Eight cards would be overkill for most of the batch size interval considered, unless the service observes heavy sustained traffic.

\subsection{Synthetic Inputs Generation}

Our benchmark framework is designed to support both synthetic and real inputs for comprehensive performance evaluation. This dual approach enables us to conduct controlled experiments while ensuring relevance to practical applications. Both input types can accommodate various data modalities including text, images, or combinations of different modalities.

Our synthetic input generation framework addresses the diverse request patterns encountered in production LLM inference deployments. By systematically sampling the request distribution space, our synthetic inputs can be precisely calibrated to target specific performance characteristics and edge cases. We focus on several key dimensions of inference request diversity:

\begin{itemize}
\item \textbf{Input Modality Distribution}: Configurable distributions across text, image, and audio modalities with adjustable parameters to evaluate performance under varying workloads
\item \textbf{Sequence Length Variability}: Parameterized length distributions that enable testing how system performance scales differently with input length ($L$) versus output length, as only prefill complexity depends on $L$
\item \textbf{Intra-batch Heterogeneity}: Varying computational complexity within batches to test diverse computational patterns
\item \textbf{Computational Patterns}: Generated inputs that specifically target distinct memory and computation behaviors—where prefill performs bulk matrix operations while decoding primarily involves cache updates
\end{itemize}

For text workloads, our generators produce either uncorrelated random batches or temporally correlated request streams with configurable correlation coefficients. For vision-language workloads, we parameterize image resolution, complexity, and text-image coupling. Our framework enables straightforward comparison between synthetic and real inputs across all these dimensions.

\begin{figure}[htbp]
    \centering
    \begin{subfigure}[b]{0.45\textwidth}
        \centering
        \includegraphics[width=\textwidth]{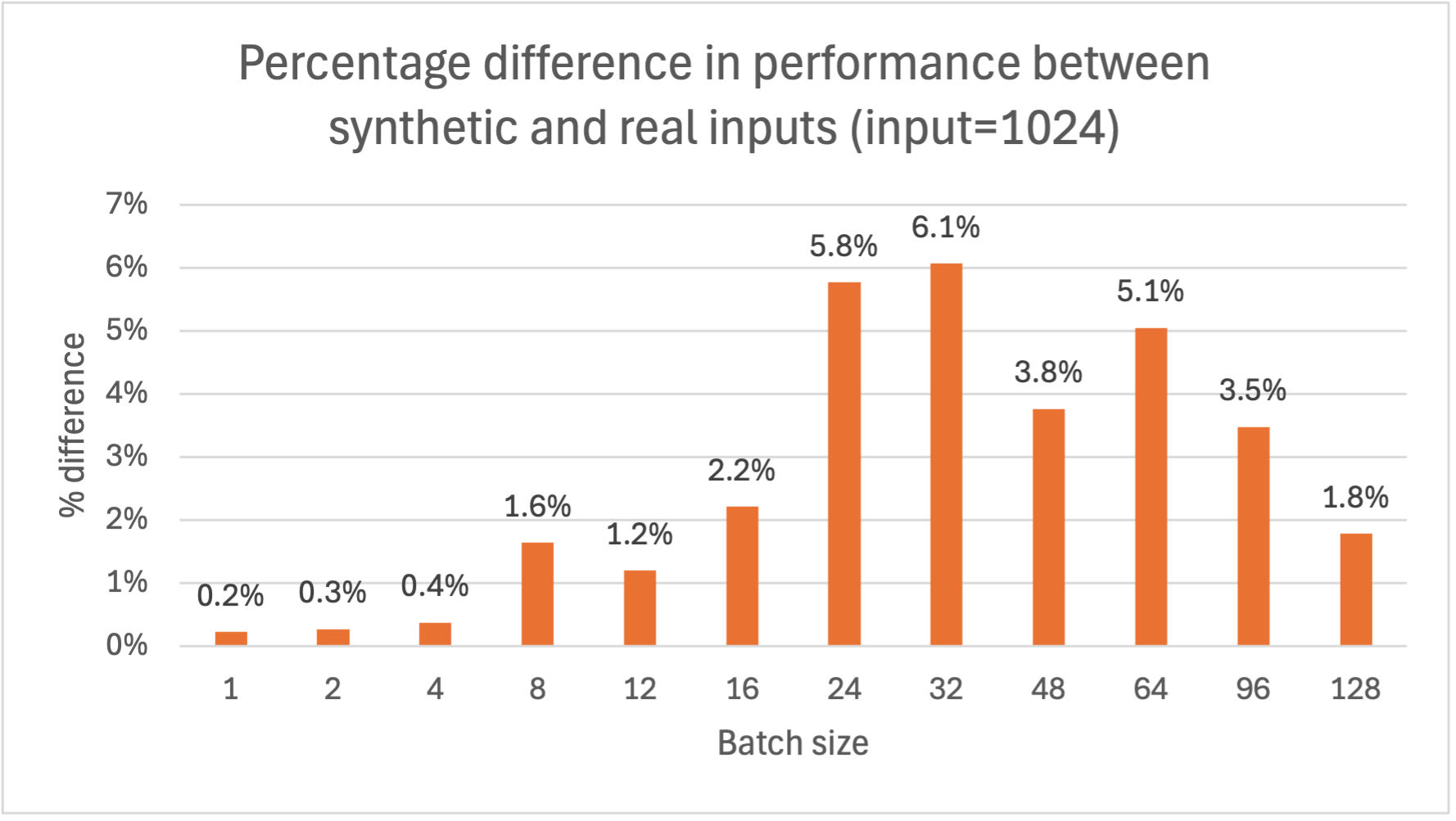} 
        \caption{Performance difference between synthetic and real inputs at sequence length 1024}
        \label{fig:synthetic-input1}
    \end{subfigure}
    \hfill
    \begin{subfigure}[b]{0.45\textwidth}
        \centering
        \includegraphics[width=\textwidth]{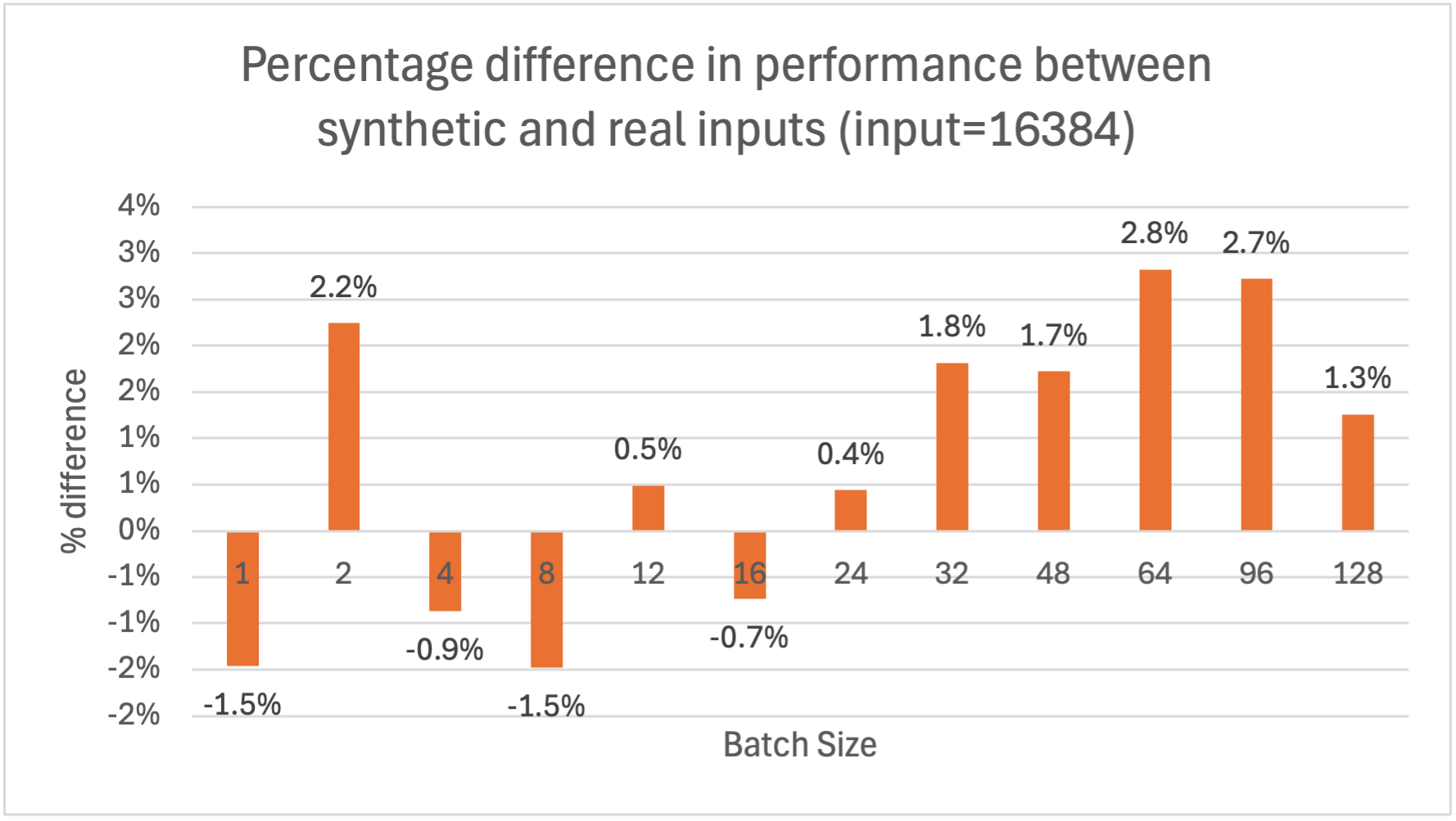} 
        \caption{Performance difference between synthetic and real inputs at sequence length 16384}
        \label{fig:synthetic-input2}
    \end{subfigure}
    \caption{Performance comparison between synthetic and real inputs at sequence lengths of 1024 and 16384 tokens across batch sizes 1-128. Values show percentage deviation, with positive values indicating synthetic inputs perform faster than real inputs.}
    \label{fig:synthetic-input}
\end{figure}

Figure~\ref{fig:synthetic-input} illustrates the percentage difference in throughput between synthetic and real inputs at two representative sequence lengths across various batch sizes. The results show minimal differences between synthetic and real inputs, with particularly tight alignment at smaller batch sizes. While only two sequence lengths are presented here, our experiments across sequence lengths from 1024 to 16384 show similar trends. Across the entire parameter space, differences consistently remain small---with a geomean percentage difference of only 2.37\%. Even in the most divergent case (batch size 96 with sequence length 8192), the difference remains below 7.01\%.

These minimal variations validate that our synthetic generators accurately approximate real workload characteristics, enabling controlled experiments without distorting results.

% \nelson{
% \textbf{[Discuss]} Models --  how model parameters affect different performance metrics; for this we developed a synthetic model generator
% }

\subsection{Synthetic Model Generation}

We designed a synthetic model generator to systematically investigate the relationship between model parameters and inference performance metrics. We build Transformers-based models with configurable dimensions and layer counts by mimicking the configuration parameters of real models. The implementation enforces architectural constraints, such as requiring hidden dimensions to be multiples of 64 to maintain compatibility with typical attention mechanism implementations. The generator preserves realistic proportionality between parameters, such as setting the intermediate size to 4x the hidden dimension and calculating appropriate attention head counts based on dimension. 

\begin{figure}[!h]
    \centering
    \includegraphics[width=0.750\linewidth]{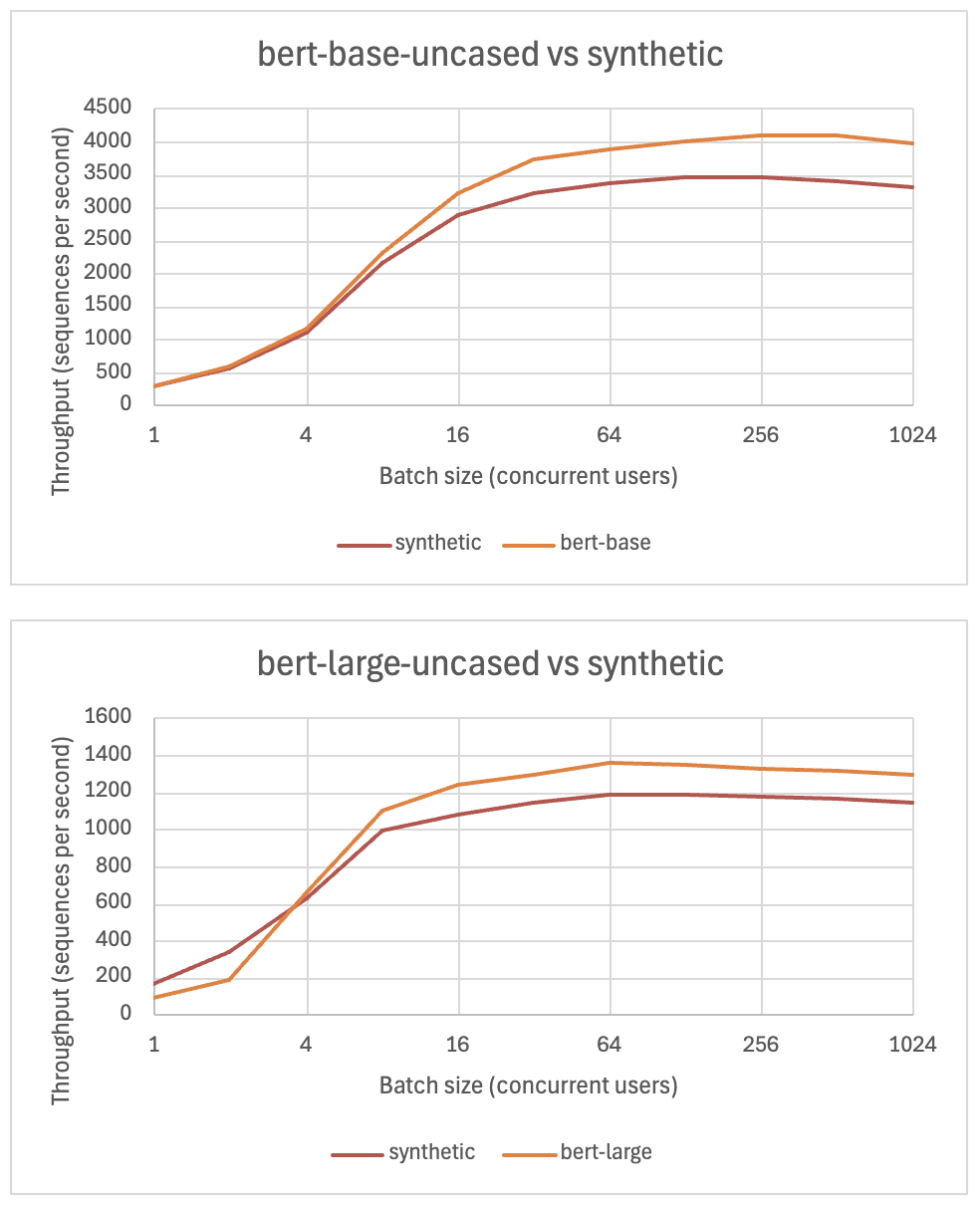}
    \caption{Throughput (in sequences processed per second) for different encoder models and their synthetic counterparts.}
    \label{fig:synmod}
\end{figure}

\begin{figure*}[t]
    \centering
    \includegraphics[width=1.000\linewidth]{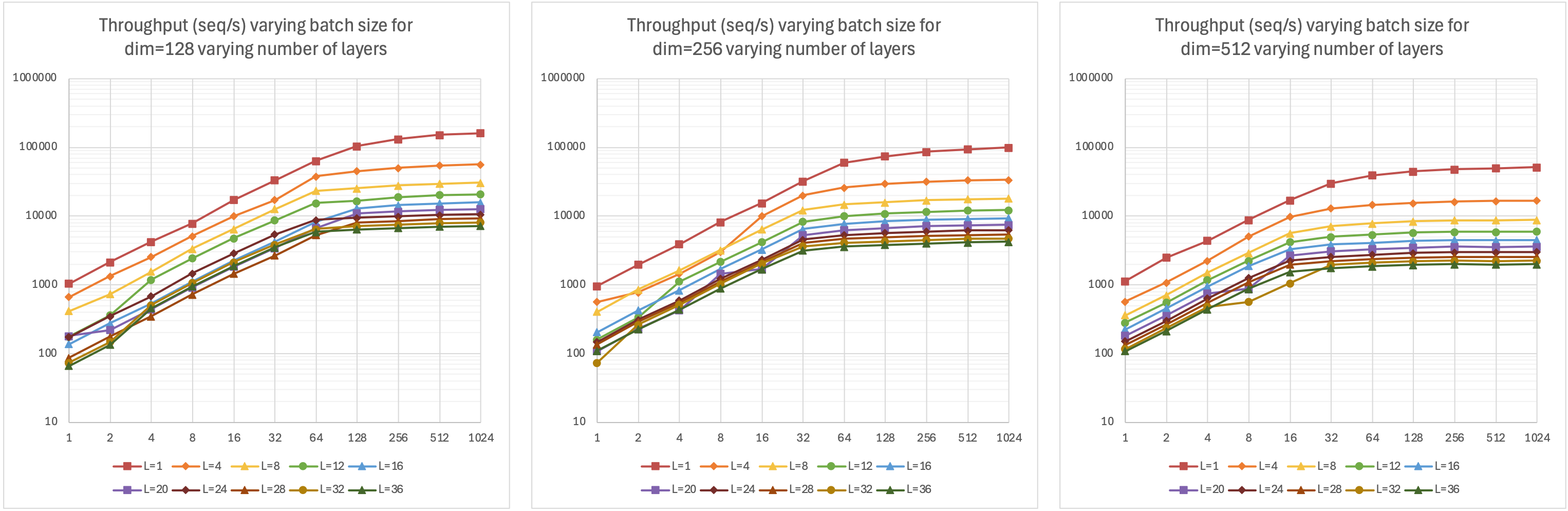}
    \caption{Throughput (in sequences processed per second) of encoder models with different geometries.}
    \label{fig:syncomp}
\end{figure*}

Figure~\ref{fig:synmod} shows performance (throughput, in input sequences processed per second) of two encoder models — bert-base~\footnote{https://huggingface.co/google-bert/bert-base-uncased} and bert-large~\footnote{https://huggingface.co/google-bert/bert-large-uncased} — and their synthetic counterparts — meaning, models dynamically built mimicking the geometry of the original model. Experiments were executed with \FMwork running a minimum number of 10 inference calls, then iterating for as long as necessary until the timing variability across iterations is under 5\%. We also determine a maximum duration to run such experiment (for encoder models, typically a minute). We calculate throughput as number of sequences processed (batch size) by median iteration time (excluding a number of initial iterations which are considered warmup; typically 20\%). Comparing the results, we see a geomean percentage difference of 10\% across batch sizes (varying from 1 to 1024 in multiples of 2). Experiments are executed with sequence length of 512 tokens.

The results of this parametric analysis reveal significant correlations between architectural decisions and performance. Figure~\ref{fig:syncomp} shows how performance varies for different values of dimensions and number of layers running experimental sweeps with input size 512 tokens varying batch sizes. We execute the sweeps varying model dimension (128 to 1024 in increments of 128) and number of layers (1 and then 4 to 32 in increments of 4). As we increase dimensionality, throughput naturally decreases. However, the difference in performance is clearer for larger batch sizes. For instance, for a model with 12 layers (e.g., bert-base), at batch size 1024 (1024 sequences processed in one inference call), a model with only 128 dimensions is almost 9x faster than a model with 1024 dimensions. The difference at batch size 4 is only 8\%. 

\begin{figure}[!ht]
    \centering
    \includegraphics[width=0.800\linewidth]{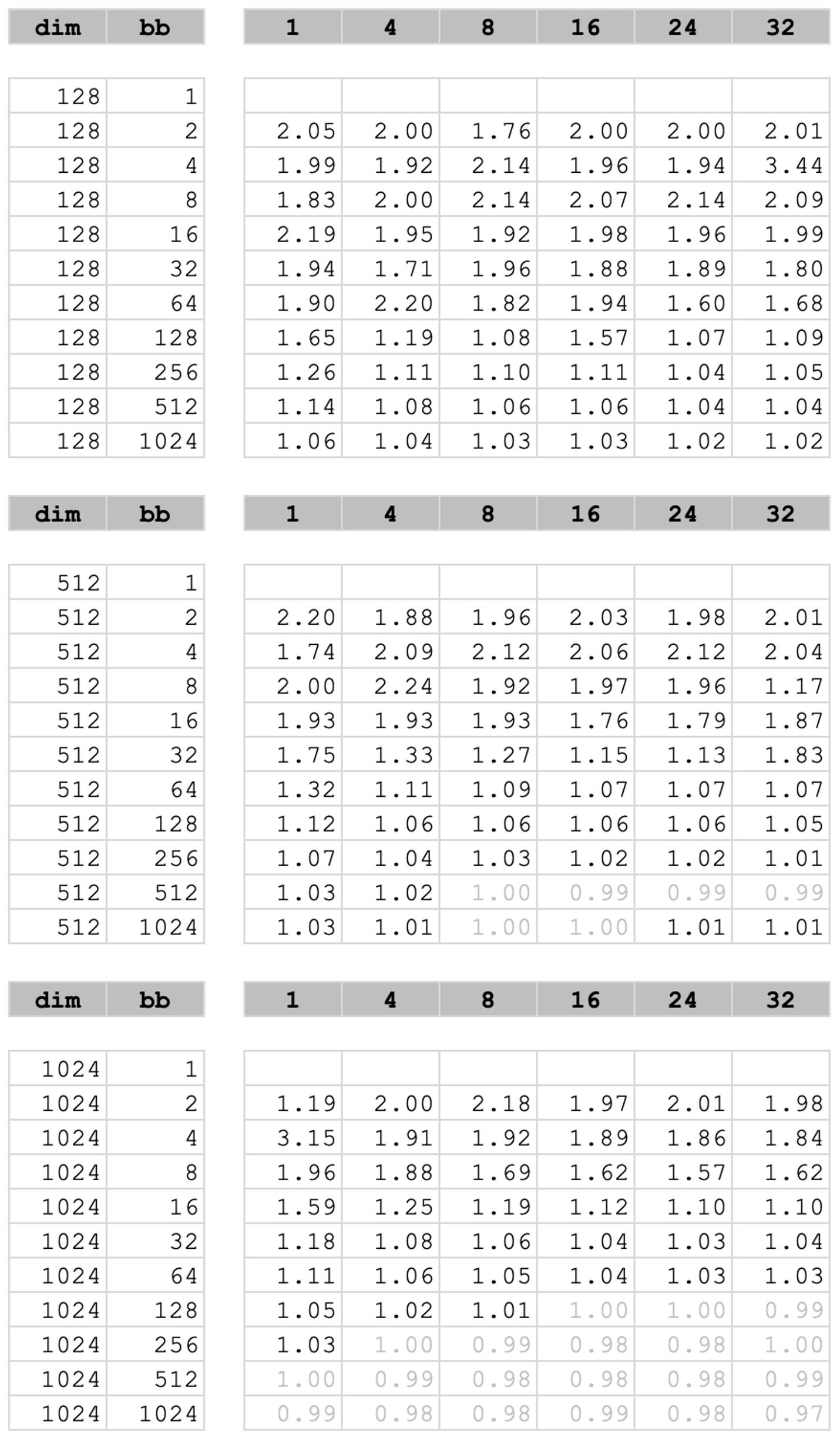}
    \caption{Incremental factor of throughput of encoder models varying geometry; dimensions: 128, 512, 1024; number of layers as columns: 1, 4, 8, 16, 24, 32.}
    \label{fig:syntab}
\end{figure}

The experimental sweeps are also useful for detecting optimal operating ranges for different models. Each combination of geometry parameters leads to a throughput progression that approaches an asymptote after a certain batch size is reached. The tables in Figure~\ref{fig:syntab} show this progression for different synthetic encoder models varying dimension (128, 512, 1024) and number of layers (columns, from 1 to 32). Each sweep varies batch size. The values in the table are throughput factors from the previous batch size to the next. Since the batch sizes are multiples of 2, the perfect factor would be 2.0. A factor of 1.0 means no improvement in throughput, and a factor below 1.0 means performance degradation. For a model with dimension 128, for instance, throughput keeps increasing beyond 5-10\% even for larger batch sizes. In contrast, for a model with dimension 1024, for most tests varying number of layers the optimal batch size is between 64 and 128; any concurrency value beyond that number does not increase throughput and, since the hardware device becomes saturated, throughput may actually decrease.

These synthetic model experiments demonstrate our ability to isolate architectural parameters and their performance impacts. By controlling model geometry while maintaining realistic proportions, we enable hardware-aware design decisions and performance predictions. This capability complements our synthetic input generation, providing comprehensive control over both model and workload characteristics in our benchmarking methodology.
\section{Related Work}
\label{sec:relatedwork}

The rapid growth of LLMs has created an urgent need for efficient inference benchmarking. Various approaches have been proposed, yet most fail to efficiently navigate the vast experimental space of LLM inference. Our work addresses this limitation through a systematic approach that significantly reduces experimental complexity while maintaining high accuracy.

% 1. Standardization and Experimental Efficiency Challenges

In the field of LLM inference benchmarking, the MLPerf Inference Benchmark Suite \cite{reddi2019mlperf} serves as an industry standard providing predefined tests to compare system performance across vendors, designed for standardized cross-platform evaluation with fixed workloads and metrics; by contrast, our \FMwork addresses a different purpose: enabling users to efficiently benchmark their own LLM configurations on specific hardware while reducing experimental complexity through meta-metrics and intelligent parameter selection strategies. LLM-Inference-Bench \cite{chitty2024llm} provides a comprehensive suite for evaluating LLM inference across diverse accelerators, models, and frameworks, offering valuable insights into performance trade-offs, but unlike our \FMwork, it relies on exhaustive testing without methodologies to reduce experimental complexity. Similarly, EchoSwift \cite{krishna2024echoswift} offers configuration discovery and performance benchmarking modes for LLM deployment, identifying optimal service configurations for specific LLM requirements, yet it lacks synthetic model generation capabilities to analyze parameter impacts efficiently and provides no quantitative methodology to balance experimental cost against accuracy.

% 2. Hardware-Model Matching Optimization Challenges

Several systems focus specifically on performance prediction and hardware recommendations for LLM deployment: LLM-Pilot \cite{lazuka2024llm} characterizes and predicts LLM inference performance across different GPUs to recommend cost-effective hardware for deployment; DeepSpeed-Inference \cite{rasley2020deepspeed} enables efficient transformer inference with excellent latency and throughput performance; Emani et al. \cite{emani2022comprehensive} conducted a comprehensive evaluation of AI accelerators across various workloads, finding that different accelerators excel at different workload types; and Vidur \cite{agrawal2024vidur} uses simulation to predict inference performance metrics without extensive hardware testing. These systems primarily optimize specific model-hardware combinations or provide hardware recommendations rather than addressing the efficiency of the benchmarking process itself; they do not incorporate experimental design principles to efficiently reduce the search space—a limitation precisely addressed by our \FMwork through structured parameter selection methodology that optimizes experimental efficiency while maintaining benchmarking reliability, providing what organizations still need: a methodology to reduce the extensive experimental overhead required for performance characterization.

% 3. Sustainability and Resource Constraint Challenges

Energy-aware LLM evaluation represents an emerging area of research that examines the environmental impact of inference: From Words to Watts \cite{samsi2023words} studies the energy consumption of LLaMA models across different GPU architectures, highlighting how power capping can improve efficiency, while recent optimization-focused works \cite{dao2023flashattention, shah2024flashattention, lin2024awq, kwon2023efficient} benchmark specific improvements in quantization, memory management, and attention computation. However, these approaches do not tackle the challenge of optimizing the benchmarking process itself which is precisely what \FMwork addresses. By leveraging meta-metrics, parameter selection, and cost-performance analysis, \FMwork enables systematic and efficient benchmarking that can be applied to both performance and energy-aware evaluations, making these approaches complementary rather than competitive.

%Through our systematic approach with meta-metrics that quantify experimental cost-accuracy trade-offs, intelligent parameter selection strategies, and synthetic model generation capabilities, we significantly improve benchmarking efficiency while maintaining high accuracy. Our framework achieves comparable accuracy with significantly fewer tests, addressing the critical need for efficient benchmarking methodologies in the expanding landscape of LLM inference. This approach enables rapid performance evaluation that complements existing hardware recommendation capabilities and energy-aware evaluations, dramatically reducing the experimental overhead required for comprehensive LLM inference benchmarking.
\section{Conclusion}
\label{sec:conclusion}

This paper presents FMWork, a structured and efficient framework to benchmarking LLM inference performance. Our framework addresses the fundamental challenge of navigating the vast experimental landscape without requiring exhaustive testing of every possible configuration. \FMwork leverages a combination of meta-metrics that quantify trade-offs between experimental cost and accuracy, strategies for selecting high-impact parameter combinations, and synthetic model and input generation to simulate diverse workloads. We demonstrate the effectiveness of \FMwork through significant improvements in experimental efficiency. Compared to evaluations using full ground truth datasets, our methodology achieves up to a 24x speedup in benchmarking effort while maintaining high accuracy. Furthermore, even when applied to already optimized subsets of the experimental space, we show an additional 2.7x gain when reducing the output size from 1024 to 128 while preserving 96.6\% of the original accuracy.

\bibliographystyle{plain}
\bibliography{references}

\end{document}